\documentclass[pre,aps,reprint,twocolumn,showpacs]{revtex4}
\usepackage{graphicx}
\usepackage{amsmath}
\usepackage{amssymb}
\usepackage{color}
\usepackage[latin2]{inputenc}
\def\slaninafigdir{.}
\begin{document}
\title{%
Localization in random bipartite graphs: numerical and empirical study
}
\author{%
Franti\v{s}ek Slanina
}
\affiliation{%
Institute of Physics,
 Academy of Sciences of the Czech Republic,
 Na~Slovance~2, CZ-18221~Praha,
Czech Republic
}
\email{
slanina@fzu.cz
}
\begin{abstract}
We investigate adjacency matrices of bipartite graphs with power-law
degree distribution. Motivation for this study is twofold. First,
vibrational states in granular matter and jammed sphere packings;
second, graphs encoding social interaction, especially electronic
commerce. We establish the position of the mobility edge, and show
that it strongly depends on the power in the degree distribution and on
the ratio of the sizes of the two parts of the bipartite graph. At
the jamming threshold, where 
the two parts have the same size, localization vanishes. We found
that the multifractal spectrum is non-trivial in the delocalized phase,
but still near the mobility edge. We also study an empirical
bipartite graph, namely the Amazon reviewer-item network. We found
that in this specific graph mobility edge disappears and we draw a
conclusion from this fact regarding earlier empirical studies of the
Amazon network.
\end{abstract}
\pacs{%
05.40.-a;
89.75.-k;
63.50.Lm}
\date{\today}%
\maketitle%
\section{Introduction}
Localization of eigenvectors is a phenomenon common to disordered
systems. Since the pioneering work of P.~W.~Anderson
\cite{anderson_58}, very large amount of knowledge was accumulated
\cite{abrahams_10,lee_ram_85,kra_mcki_93,eve_mir_08},  yet 
rigorous answers are scarce \cite{ku_sou_83,stollmann_01}. Mostly we
must rely on qualitative approaches \cite{abr_and_lic_ram_79},
analytical approximations based on diagrammatic methods
\cite{vol_wol_80,suslov_95,jan_kol_05}, replica trick
\cite{wegner_79a} or supersymmetry \cite{efetov_83}, often using Bethe
lattice as simplified geometry
\cite{ab_an_tho_73,ab_tho_74,log_wol_85,ant_eco_77,gir_jon_80,efetov_90,mi_fyo_91}. 
Often the ``brute force'' numerical approaches lead to most reliable
answers \cite{markos_06,mon_gar_10}.

Here we look at localization on random graphs \cite{bollobas_85}.
We investigate the eigenvalues and eigenvectors of the adjacency
matrix, which encodes the structure of the graph. Therefore, the
disorder is purely off-diagonal, in contrast with e. g. the model of a quantum particle in 
 random potential.

There are numerous motivations for studying  spectra and localization in 
random graphs. Obviously, topologically disordered materials, like
glasses, are common and investigation of their electronic and vibrational
spectra has high practical relevance. Random graphs are a natural choice
for modeling these structures.
{In the tight-binding approximation, the
Hamiltonian of an electron in such structure composed of atoms of the
same type (like a metallic glass) is proportional to the adjacency
matrix of the graph. Hence the motivation for the study of spectral
properties of adjacency matrices of random graphs.}

{As another example,} granular materials
\cite{ja_na_be_96a} exhibit highly complex distribution of internal
stress, often described in terms of force chains (see
e. g. \cite{maj_beh_05}). Sound propagates 
mainly along these chains \cite{ji_ca_ve_99,owe_dan_11,bas_owe_dan_por_12}, so 
  we can make an abstract model of a  
granular matter in terms of a (random) graph representing the force
chains.
{Vibrational states of the granular material then correspond to
eigenstates of the Laplacian defined on the graph.} Unusual behavior of
low-energy vibrations 
in non-crystalline solids leads to anomalous thermal conductivity in
such materials \cite{zel_poh_71}, which founds close analogy also in
granular materials \cite{xu_vit_wya_liu_nag_09}.


Interestingly, the physics of glasses and granulars found recently a
common ground in terms of the jamming transition
\cite{ca_wi_bou_cla_98b,liu_nag_10,liu_nag_saa_wya_11,tor_sti_10}.
Packing of hard spheres is an extremely complex problem with
ramifications in various disciplines
\cite{ast_wea_08,con_slo_99}. Jamming transition occurs when average
number of contacts is just sufficient for mechanical stability. 
Recently a model was proposed
\cite{parisi_14,fra_par_16,fra_par_urb_zam_16} which relates sphere
packing to combinatorial optimization \cite{nishimori_01}. In a
typical setting, $N$ objects must satisfy $M$ constraints. This may be
formulated as minimization problem for a Hamiltonian of $N$ variables,
composed from $M$ additive terms. In a graph-theoretic language, the
problem can be formulated in terms of a bipartite graph, with a set of $N$
variables on one side and set of $M$ constraints on the other
side. When the Hamiltonian is expanded to harmonic approximation, its
eigenmodes are related to the eigenvectors of the underlying bipartite
graph. Hence the importance of studying bipartite graphs for the
jamming problem.
 
{Of course, the approach of
Refs. \cite{parisi_14,fra_par_16,fra_par_urb_zam_16}  is distant from
real systems in the sense that they construct a kind of mean-field
jamming transition, in which the number of contacts between
spheres goes to infinity. However, such methodology proved already
useful many times, especially in the theory of spin glasses
\cite{nishimori_01}, which justifies its use also for the jamming
problem. Formally it is manifested by replacing the real graph of
contacts, which is embedded into three-dimensional Euclidean space, by
a random graph which is effectively infinite-dimensional. We believe
this also justifies the use of jamming terminology in the case of the 
random graphs used in this work. At the same time, we should keep in
mind that graphs pertaining to realistic models of jamming should have
quite narrow 
degree distribution. Therefore, in the context of our scale-free
graphs we should rather speak of ``abstract'' of ``generalized'' jamming
problem. In this sense we can speak of jamming transition in any
bipartite graph.}



{
However, our immediate motivation comes from the study of 
 bipartite random graphs that naturally occur
in electronic commerce. They belong to a broader class of scale-free
graphs (i. e. those with power-law degree distribution)
\cite{alb_bar_01}. At this point let us make just brief remark that related
problems were also investigated in the field of correlation matrices
\cite{la_ci_bou_po_99,ple_gop_ros_am_sta_99,ciz_bou_94,bur_jur_now_04}.
}

We already studied several electronic-commerce 
networks in the past \cite{sla_kon_10,slanina_12a,slanina_14}. Here we
shall reexamine the Amazon network  \cite{sla_kon_10}. It is a bipartite graph of
reviewers on one side and items offered for sale on the other side. We
found that the degree distribution follows a power law on both
sides. Moreover, we found by diagonalization of the corresponding
matrix that the localized eigenvectors carry  non-trivial semantic
information on the network. {
Indeed, we were able to clearly identify several
groups of agents sharing the same interests. Therefore, we found a
practical application of the study of localization in empirical
networks. However, it would be highly desirable to have a model of
such network, at least to provide certain benchmark as to density of
eigenvalues and dependence of the inverse participation ration on
eigenvalue. We propose a random bipartite graph with power-law degree
distribution as a model of these empirical networks. Here we want to
study how much the model reproduces the empirical data as to spectrum
and localization properties.}

 For completeness we should also mention that localization
was already used in extracting information from scale-free graphs,
e. g. in
Refs. \cite{sad_kal_hav_ber_05,zhu_yan_yin_li_08,jal_sol_vat_li_10,gir_geo_she_09,odor_14}. 

So, the aim of this work is investigation of spectra and especially
the localization in bipartite random graphs with power-law degree
distribution (usually called scale-free graphs, although this term is
somewhat misleading). A good deal of information was already obtained on the
spectra of scale-free graphs. To cite just a few articles, see
\cite{far_der_bar_vic_01,goh_kah_kim_01b,dor_gol_men_sam_03}. The most
important finding is that the power-law degree distribution induces a
power-law tail in the density of eigenvalues. This is found generally,
irrespective of the specific model used for the scale-free graph.

From the mathematical point of view, spectra of random graphs are just
spectra of a special type of random sparse matrices. 
Analytical approaches exist for 
the density of eigenvalues, using the replica trick
\cite{rod_bra_88,sem_cug_02,rod_aus_kah_kim_05,nag_rod_08} or cavity 
approach \cite{ciz_bou_94,cav_gia_par_99,kuhn_08} (we proved that these two methods
are 
strictly equivalent in \cite{slanina_11}), or, alternatively, by mapping on a
supersymmetric Hamiltonian \cite{rod_ded_90,fyo_mir_91,fyo_mir_91a}.

Both replica trick and cavity method provide a ground for systematic
analytic approximations, like effective-medium and single-defect
approximations,  which grasp essential features of the
spectrum, like in Refs.  \cite{rod_bra_88,bi_mo_99,sem_cug_02,slanina_11}.  
  Of course, the cavity equations can
be also solved by brute force using numerical population algorithms
\cite{ciz_bou_94,cav_gia_par_99,cil_gri_mar_par_ver_05,kuhn_08,rod_cas_kuh_tak_08,met_ner_bol_10,bir_sem_tar_10,mon_gar_11,kuh_mou_11,kuhn_16}. 

On
the contrary, it is much harder to find similar analytical approximations to describe
localization.  Some of the difficulties encountered in
attempts to find analytical approximations were investigated by us
earlier \cite{slanina_12b}. Therefore, numerical solution of the
cavity equations is usually used as a reliable method
\cite{ab_an_tho_73,ab_tho_74,cav_gia_par_99,cil_gri_mar_par_ver_05,met_ner_bol_10}. 
Of course, it is always 
possible to pursue the study by
direct numerical diagonalization of 
sample random graphs
\cite{evangelou_83,evangelou_92,eva_eco_92,ciz_bou_94,bi_mo_99,bir_sem_tar_10,mon_gar_11,slanina_12b,bir_tei_tar_12,tik_mir_skv_16}. 
We shall resort to
the 
latter approach here. 

In our previous work we investigated localization on Erd\H{o}s-R\'enyi
random graphs and on random regular graphs \cite{slanina_12b}. As
we already hinted above, here we turn to localization on bipartite
graphs.  Spectra of such graphs, maybe under various disguises, were
already studied earlier \cite{nag_tan_07,kho_rod_97,slanina_11}. We
took inspiration from the work \cite{nagao_13}, where the spectrum of
a scale-free bipartite graph was studied by replica approach within the
effective-medium approximation. The structure of the graphs we shall
construct will follow the algorithm of Goh et
al. \cite{goh_kah_kim_01a}. This is a natural generalization of the
Erd\H{o}s-R\'enyi graph ensemble to the case of non-uniform
probabilities of placing edges between vertices. Spectra of these
graphs were studied in Ref. \cite{rod_aus_kah_kim_05} and the
generalization to bipartite graphs was investigated in the already
mentioned Ref. \cite{nagao_13}. Our immediate aim is to add the
aspect of localization to these studies.

\section{Scale-free bipartite graph}

\subsection{{Relevance of the adjacency matrix}}

{We shall deal with localization due to topological disorder, rather
than random on-site  potential. In the language of an electron moving in a
random lattice represented by a graph $G$, we start with a general
tight-binding Hamiltonian
\begin{equation}
H=\sum_i\epsilon_i|i\rangle\langle i|+\sum_{i<j}t_{ij}(|i\rangle\langle j|+|j\rangle\langle i|)
\end{equation}
with on-site energies $\epsilon_i$ and hopping terms $t_{ij}$
connecting vertices $i$ and $j$ of the graph. 
Then we make a special choice relevant to our case, namely all diagonal
elements equal (and without loss of generality they may be all zero)
and hopping terms having value $t_{ij}=t$ if $(i,j)$ is an edge in the graph
$G$ and $t_{ij}=0$ otherwise. Such situation occurs e. g. in metallic
glasses. Indeed, all atoms are equal but the local structure may change from
one site to another. Then, the Hamiltonian of the particle is
proportional to the adjacency matrix of the graph $G$. Again, setting
the proportionality constant $t=1$ just fixes the energy
scale. Therefore, all essential information is obtained in the
spectrum and eigenvectors of the adjacency matrix of the graph.}

{On the other hand, in the study of vibration states of glasses and
granular matter we  need to diagonalize the matrix representing the
Laplacian on the graph. However, when we study the localization on
random graphs, there is a disadvantage in using the Laplacian. Indeed,
we want to separate the effect of on-site disorder (random atomic
energy in tight-binding electronic Hamiltonian or random atomic mass
in a model of vibrations) from the effect of random graph topology. In
the Laplacian, this two effects are mixed, because diagonal element is
related to the degree of the node in the graph. This aspect makes the analysis
less transparent. Therefore, we prefer to study the spectrum of
adjacency matrix, where on-site disorder is totally absent.}

{The third reason for studying adjacency matrix 
lies in our previous empirical study of Amazon network
\cite{sla_kon_10}, which was done 
using adjacency matrix. Small communities in the network were
successfully found by studying localization. To make comparison with a
model random graph, adjacency matrix is studied also here.}

\subsection{Algorithm for graph creation}

Now let us turn to our specific type of bipartite graph. 
The algorithm for creating instances of our random graph follows the
original idea of Goh et al. \cite{goh_kah_kim_01a}, further adapted by
Nagao \cite{nagao_13}. We have two sets of vertices, the set A containing
$N$, the set B containing $M$ vertices. We shall assume $N\le M$. 
 Among these vertices, $L$ edges
are distributed, connecting always a vertex from A to a vertex from
B. This way, a bipartite graph is created.  The vertices are not 
statistically equivalent. The vertex $i$ from A is given an a priori probability
$P_{Ai}$, similarly the vertices from $B$ will have probabilities
$P_{Bj}$. To construct a scale-free graph, the probabilities will have
the following power-law form
\begin{equation}
\begin{split}
P_{Ai}&=\frac{i^{-\alpha_A}}{\sum_{l=1}^Nl^{-\alpha_A}}\;,\;\;i=1,2,\ldots,N\\
P_{Bj}&=\frac{j^{-\alpha_B}}{\sum_{l=1}^Ml^{-\alpha_B}}\;,\;\;j=1,2,\ldots,M\;.
\end{split}
\end{equation}
The edges are placed in the following way. In each step, a pair of
vertices $(i,j)$ from A and B, respectively,
 is chosen randomly with probability $P_{Ai}P_{Bj}$. 
If an edge connecting $i$ and $j$ already exists, the
choice is canceled and a new pair is randomly selected. (This may be
repeated several times, if the graph is already rather dense, but as
long as
{$L \le N M$}, a pair is eventually found.)
Otherwise, a new edge is placed connecting $i$ and $j$. 
Repeating this procedure $L$ times we obtain a graph with exactly $L$
edges and there are no multiple edges. It was found that the
cumulative degree distribution on the A side has a power-law tail 
$P^>(k)\sim k^{-\gamma}$, where $\gamma=1/\alpha_A$ for $\alpha_A<1$,
while $\gamma=1$ for all $\alpha_A\ge 1$ 
\cite{goh_kah_kim_01a,lee_goh_kah_kim_04,fle_sok_13}. By symmetry,
analogous formulas hold for the degree distribution on the B side.

The structure of the graph is encoded in the adjacency matrix, which
has, due to the bipartite character, the following form
\begin{equation}
R=\left(
\begin{array}{cc}
0 & S\\
S^T & 0
\end{array}
\right)
\end{equation}
where $S$ is an $N\times M$ rectangular matrix. The spectrum of the
matrix $R$ was studied using the replica method in
Ref. \cite{nagao_13}. It was found that the density of eigenvalues has
a power law tail  which depends only on the greater of the two exponents
$\alpha_A$, and $\alpha_B$. This suggests that it is sufficient to
study graphs with both exponents equal, $\alpha_A=\alpha_B=\alpha$,
which is what we shall assume in the following. 

The matrix $R$ has size $(N+M)\times(N+M)$, which can be huge, as will
be the case e. g. in the empirical 
data studied in the last section of this paper.  
However, essentially the same information on the spectrum and
eigenvectors can be obtained from diagonalization of a smaller matrix
$C=S\,S^T$ of size $N\times N$. Obviously, if $e$ is an eigenvector of $C$ with
eigenvalue $\lambda^2$, then $\left(\begin{array}{c}
e\\
S^Te/\lambda
\end{array}
\right)$
is an eigenvector
of $R$ with eigenvalue $\lambda$
{(see Appendix B, if
unclear). Choosing plus or minus sign of
$\lambda$ we find that single eigenvector $e$ of matrix $C$
corresponds to just two independent eigenvectors of $R$.}
If we are interested only in 
localization on the A side, knowledge of eigenvectors of $C$ is
just sufficient.
{If we needed also elements of the eigenvector on the
B side, they can be reconstructed from the eigenvector $e$ of matrix
$C$.} {Therefore, all computations in this article are for the
matrix $C$.}

 Using the replica method, it was found that the power-law
tail of the density of eigenvalues of the matrix $C$ is $\mathcal{D}(z)\sim
z^{-1-\tau}$, where $\tau=1/\alpha$ \cite{nagao_13}.  Note that the
exponent for the density of eigenvalues is the same as the exponent 
for the degree distribution.

\begin{figure}[t]
\includegraphics[scale=0.85]{%
\slaninafigdir/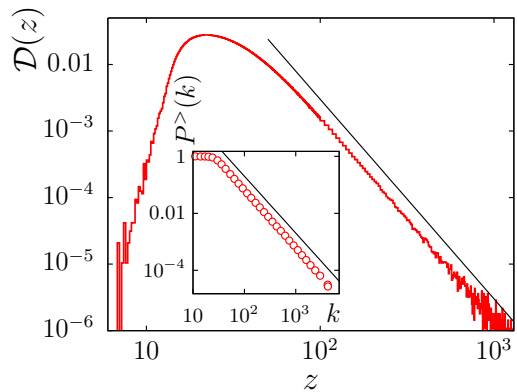}
\caption{Density of eigenvalues for the graph with $\alpha=1/2$,
  $M/N=18$, $L/N=50$,  and size $N=30000$. The straight line is the power $\propto
  z^{-3}$. In the inset, the cumulative degree distribution in the A
  set, for the
  same graph. The
  straight line is the power $\propto k^{-2}$.
}
\label{fig:randbip-dos-0.5-18-50}
\end{figure}

\begin{figure}[t]
\includegraphics[scale=0.85]{%
\slaninafigdir/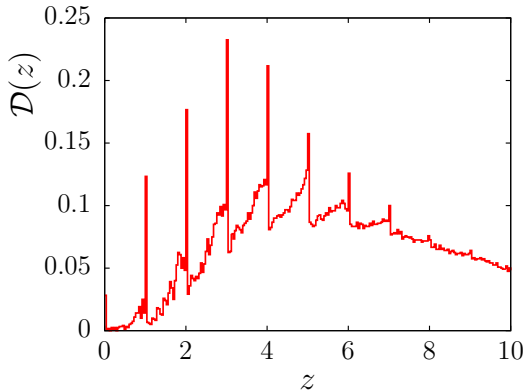}
\caption{Detail of the density of eigenvalues for the graph with $\alpha=1/2$,
  $M/N=18$, $L/N=10$. 
}
\label{fig:randbip-dos-0.5-18-10}
\end{figure}

\begin{figure}[t]
\includegraphics[scale=0.85]{%
\slaninafigdir/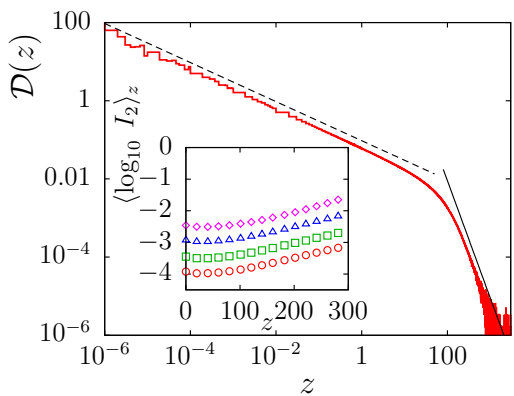}
\caption{Density of eigenvalues for the graph with $\alpha=1/2$,
 $N=30000$, $M/N=1$, $L/N=50$. The straight lines are the powers $\propto
  z^{-1/2}$ (dashed) and $\propto z^{-3}$ (solid). In the inset,
  averaged logarithm of the inverse participation ratio for the graph
  with the same values of $\alpha$, $M/N$ and $L/N$ and sizes $N=30000$
  ({\Large $\circ$}),
  $N=10000$ ($\Box$), $N=3000$ ($\triangle$), and $N=1000$ ({\Large $\diamond$}). 
}
\label{fig:randbip-dos-0.5-1-50}
\end{figure}

\section{Spectrum and localization}

\subsection{Sample preparation}

Size dependence is the key to studying localization. Therefore, we
create artificial sample graphs of four sizes, $N=10^3$, $N=3\cdot 10^3$,
$N=10^4$, and $N=3\cdot 10^4$. For each size and each set of parameters
$\alpha$, $M/N$, and $L/N$ we created certain number of independent
samples of the random graph using the algorithm described in the previous
section. The typical number of samples was about $2.5\cdot 10^5$, $2\cdot
10^4$, $10^3$, and $50$ for $N=10^3$, $N=3\cdot 10^3$, $N=10^4$, and
$N=3\cdot 10^4$, respectively. { For each sample, the matrix $C$ was
diagonalized using the standard MATLAB library.} The first thing we checked was the
degree distribution of the graphs produced. We can see in the inset of
 Fig. \ref{fig:randbip-dos-0.5-18-50} that our algorithm created
 a graph in full agreement with analytical predictions.

\subsection{Density of eigenvalues}

We show in Fig.  \ref{fig:randbip-dos-0.5-18-50} a typical example of
the density of eigenvalues. The tail is characterized by a power-law
decay, which again, as with the degree distribution, agrees very well
with the analytic prediction. At the lower edge of the spectrum, the
density of states falls off quickly and this is the region where we
expect localization to occur. We can see that the density of
eigenvalues is rather smooth there. This is typical for large enough
$L/N$, i. e. for dense enough graphs. When the graph goes sparser,
singularities accompanied by apparent delta-functions appear at
integer eigenvalues, as can be seen in
Fig. \ref{fig:randbip-dos-0.5-18-10}.
{The quality of the data do not
allow to establish the form of the singularities. In Erd\H os-R\'enyi
graphs it was found that the singularity at the center of the spectrum
is logarithmic \cite{slanina_11}.So, by analogy we expect the
singularities have logarithmic form also here.}
We do not know of any analytical
theory which would describe this system of singularities sufficiently
well. However, the mere presence of delta-functions can be understood
by a simple consideration. Indeed, if $L/M<1$, as is the case in Fig.
\ref{fig:randbip-dos-0.5-18-10}, a macroscopic fraction of vertices in
the B set has degree $1$, i. e. does not provide a path from one
A vertex  to another one. This leads to creation of star-like
components of the graph, where a single A vertex is linked to
$m>0$ vertices from the set B, who themselves are not linked
elsewhere. Such component contributes to the spectrum of the matrix $C$
by integer value $m$.  The weight of thus created delta-function reflects the
probability with which these stars appear in the bipartite graph. This
mechanism is analogous to the appearance of delta-functions in the
spectrum of Erd\H{o}s-R\'enyi graphs, as shown numerically e. g. in
\cite{evangelou_92,eva_eco_92,kuhn_08,kuh_mou_11,slanina_11} and
analytically in \cite{bau_gol_01,golinelli_03}.

\begin{figure}[t]
\includegraphics[scale=0.85]{%
\slaninafigdir/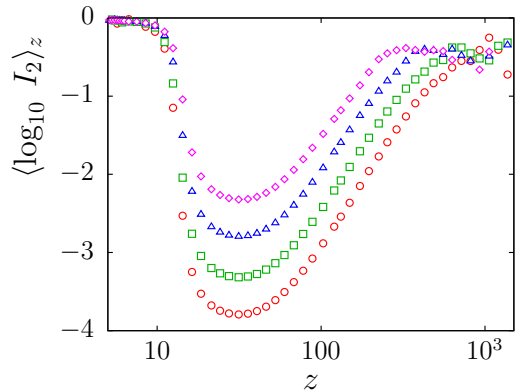}
\caption{Averaged logarithm of the inverse participation ratio for the graph
with $\alpha=1/2$, $M/N=18$ and $L/N=50$ and sizes $N=30000$ ({\Large $\circ$}),
  $N=10000$ ($\Box$), $N=3000$ ($\triangle$), and $N=1000$ ({\Large $\diamond$}). 
}
\label{fig:randbip-ipr-0.5-18-50}
\end{figure}

For all $M/N>1$ the spectrum preserves the same overall character:
there is a power-law tail at large $z$, with a power which depends
only on $\alpha$, there is a bulk of the spectrum at intermediate $z$
and an area with low density of eigenvalues at small $z$. The latter
area shrinks as $M/N$ approaches the critical value $M/N=1$, where
another power-law dependence develops. {We found that for $z\to 0$,
the density of eigenvalues exhibits a singularity
$\mathcal{D}(z)\sim z^{-1/2}$ for $M/N=1$, independently of the other
parameters $L/N$ and $\alpha$. This is demonstrated in
Fig. \ref{fig:randbip-dos-0.5-1-50}. } In fact, this is exactly the behavior
predicted for the case $M=N$ by the Mar\v{c}enko-Pastur
formula \cite{mar_pas_67}, which holds for $\alpha=0$ and $L/N\to\infty$. 

In the interpretation of
Refs. \cite{parisi_14,fra_par_16,fra_par_urb_zam_16} it corresponds to 
the critical point in the jamming transition. The $z^{-1/2}$
singularity translates into a flat density of vibrational states in
jammed granular matter, as observed numerically
\cite{her_sil_liu_nag_03,sil_liu_nag_05} as well as experimentally
\cite{bri_dau_bir_bou_10,owe_dan_13}. Our 
result implies, that the singularity at the jamming threshold is
universal, and holds for a broad range of random bipartite graphs.

\begin{figure}[t]
\includegraphics[scale=0.85]{%
\slaninafigdir/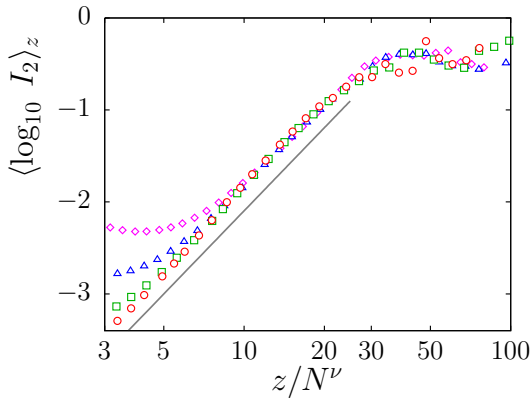}
\caption{{Detail of the averaged logarithm of the inverse participation
  ratio, versus the rescaled eigenvalue $z/N^\nu$, with $\nu=0.3$.
  The graph parameters are
 $\alpha=1/2$, $M/N=18$ and $L/N=50$ and sizes $N=30000$ ({\Large $\circ$}),
  $N=10000$ ($\Box$), $N=3000$ ($\triangle$), and $N=1000$ ({\Large
    $\diamond$}). The straight line is the dependence $3\log_{10}(z/N^\nu)+\mathrm{const}$
}}
\label{fig:randbip-ipr-0.5-18-50-righttail}
\end{figure}

\subsection{Localization}

As an indicator of localization we calculate the inverse participation
ratio (IPR), defined as $I_2(z)=\sum_{i=1}^Ne_{iz}^4$ for the eigenvector
$e_{iz}$ corresponding to the eigenvalue $z$, { normalized as
$I_1(z)=\sum_{i=1}^Ne_{iz}^2=1$ for all $z$.} Localization is revealed  
in the behavior of IPR with increasing $N$, we therefore average the
values of IPR for eigenvalues lying within an interval $(z_-,z_+)$ centered around
$z=(z_-+z_+)/2$. Numerically it is more convenient to average the
logarithm of IPR, instead of IPR itself, although we suppose that at
large enough $N$ both ways of averaging should lead to identical
conclusions about localization. Therefore, we calculate the quantity
\begin{equation}
\langle\log_{10}\,I_2\rangle_z=
\frac{1}{N_z}\sum_{z'\in(z_-,z_+)}\log_{10}\,I_q(z')
\end{equation}
where   $N_z$ is the number of eigenvalues inside the interval
$(z_-,z_+)$. We use decadic logarithm for convenience. Localized and
delocalized states differ in the dependence on the graph 
size for large $N$. Thus we have $\langle\log_{10}\,I_2\rangle_z\simeq c_0$ for $z$ in
the region of localized states, while
$\langle\log_{10}\,I_2\rangle_z\simeq c_1-\log_{10}N$ for $z$ in the
region of delocalized states. Here $c_0$ and $c_1$ are constants
independent of $N$.

{
The mobility edge $z_\mathrm{mob}$, i. e. the
value of $z$ separating localized states on one side from delocalized
ones on the other side, is extracted from the data by a procedure
described in detail later.}

\begin{figure}[t]
\includegraphics[scale=0.85]{%
\slaninafigdir/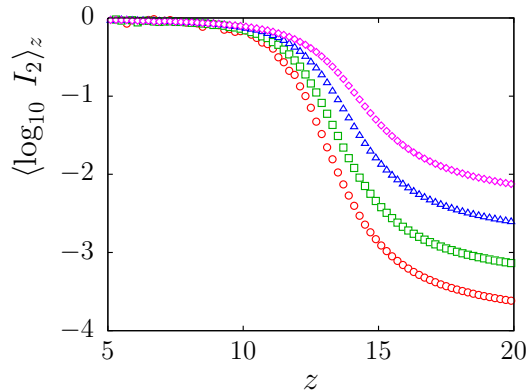}
\caption{Detail of the averaged logarithm of the inverse participation ratio for the graph
with $\alpha=1/2$, $M/N=18$ and $L/N=50$ and sizes $N=30000$ ({\Large $\circ$}),
  $N=10000$ ($\Box$), $N=3000$ ($\triangle$), and $N=1000$ ({\Large
  $\diamond$}). 
}
\label{fig:randbip-ipr-0.5-18-50-detail}
\end{figure}

We show in Fig. \ref{fig:randbip-ipr-0.5-18-50} typical behavior of
the averaged log IPR for $\alpha>0$. {We can see that the tail of the
spectrum does not exhibit a clearly defined localized regime, contrary
to the situation in Erd\H os-R\'enyi or random regular graphs
\cite{slanina_12b}. In the data, there is no clear mobility edge
visible. Instead, the region of high IPR seems to shift farther in the
tail when the graph size increases. In order to quantify this shift, we
replotted the averaged log IPR in the rescaled variable $z/N^\nu$. We
found that the best data collapse is achieved for the value of the
exponent $\nu=0.3$. The rescaled plot is shown in Fig.
\ref{fig:randbip-ipr-0.5-18-50-righttail}. The observed data collapse
suggest that in the tail the behavior of the inverse participation
ratio is
\begin{equation}
\langle \log_{10} I_2\rangle_z=\Phi(z/N^\nu)\;.
\label{eq:scalingphi}
\end{equation}
The scaling function $\Phi(x)$ exhibits two regimes, separated by a
crossover at about $x_\mathrm{cross}\simeq 30$. For $x\gtrsim
x_\mathrm{cross}$ the scaling function approaches a constant, $\Phi(x)\simeq C_> $, while
for $x\lesssim x_\mathrm{cross}$ it approaches the function $\Phi(x)\simeq
\mu\log_{10}(x)+C_< $ with coefficient $\mu=3$. The exact values
of the constants $C_<$ and $C_>$ are irrelevant, but what counts are
the values of the parameters $\mu$ and $\nu$. In fact, the observed
scaling implies the behavior $\langle I_2\rangle\sim
(z/N^\nu)^\mu$. For fixed $z$ in the tail, but within the range
$z\lesssim x_\mathrm{cross}N^\nu$ we have the dependence $\langle
I_2\rangle\sim N^{-\nu\mu}$. The value of the product $\mu\nu$ is
close but not quite equal to $1$, the exponent characteristic of
extended states. It is not clear from the available data, whether the
small difference is significant or it is due to statistical noise or
it is a finite-size effect. We consider probable that the correct
value of the product $\mu\nu$ is indeed $1$ but currently we are not
able to prove it. At present stage we can formulate a hypothesis that all the
states in the tail are extended for $z\lesssim
x_\mathrm{cross}N^\nu$. This would mean that there is no mobility edge at
the upper tail of the spectrum. However, the final verdict must be left for future.
}

The fact that localization occurs at small $z$ but, {
  as it seems,}  does not appear at 
large $z$ is in contrast with the behavior of random correlation
matrices, which in our language correspond to the value
$\alpha=0$. In this case the probabilities $P_{Ai}$, $P_{Bj}$ are
uniform, degree distribution of the graph is
Poisson, the tail of the density of eigenvalues is
steeper than any power and there is localized regime in the tail. Such
graphs therefore exhibit two mobility edges, while for $\alpha>0$ the
upper mobility edge vanishes, or, as we conjecture, is pushed far
to infinity.

\begin{figure}[t]
\includegraphics[scale=0.85]{%
\slaninafigdir/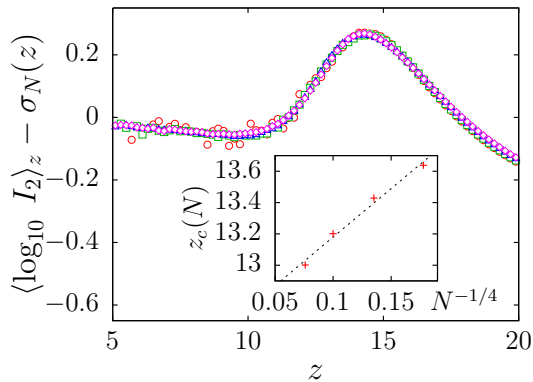}
\caption{{Data collapse of the averaged logarithm of the inverse
  participation ratio using the formulae (\ref{eq:iprcollapse}) and
  (\ref{eq:sigmoid}). The data used are identical to those in
  Fig. \ref{fig:randbip-ipr-0.5-18-50-detail}. The meaning of the
  symbols is also the same. In the 
  inset, extrapolation of the size-dependent estimates $z_c(N)$ to
  infinite size. This way the position of the mobility edge
  $z_\mathrm{mob}$ is established.} 
}
\label{fig:randbip-ipr-0.5-18-50-detail-scaling}
\end{figure}

On the opposite side, for
small $z$, {the situation is much more clear-cut, as we observe
unambiguous signs of localization.} This fact is 
sufficiently evident in the detail shown in
Fig. \ref{fig:randbip-ipr-0.5-18-50-detail}. For $z$ lower than about
$9$ the value of log IPR does not depend on graph size.
{However, such
estimate by bare inspection is not reliable enough. Let us now
describe more sophisticated procedure for establishing the mobility
edge $z_\mathrm{mob}$.}

{Below the mobility edge IPR scales with graph size as $\sim N^0$, while
above the mobility edge the behavior is $\sim N^{-1}$. However, at
finite $N$ the transition region has finite width in the variable
$z$. This suggests the following scaling for the logarithm of IPR
\begin{equation}
  \langle\log_{10}\,I_2\rangle_z=A(z)+\sigma_N(z)
  \label{eq:iprcollapse}
\end{equation}
where we denoted
\begin{equation}
  \sigma_N(z)=\sigma\Big(\frac{z-z_c(N)}{w_c(N)}\Big)\,\log_{10}N\;.
  \label{eq:sigmoid}
\end{equation}
In this expression $A(z)$ is a smooth function independent of $N$ and
$\sigma(x)$ is a sigmoid-like function with asymptotic values
$\sigma(x)\to 0$ for $x\to - \infty$ and $\sigma(x)\to -1$ for
$x\to\infty$. The size-dependent parameters $z_c(N)$ and $w_c(N)$ are
estimates of the position of the mobility edge and the width of the
transition region for given graph size. The strategy for finding the
mobility edge is to choose the sigmoid function and the set of
parameters  $z_c(N)$ and $w_c(N)$ so that the quantity
$\langle\log_{10}\,I_2\rangle_z-\sigma_N(z)$ shows best data collapse
for each four graphs sizes $N$ studied. We found that the precise
shape of the sigmoid function is not crucial. Therefore, we used the
simplest choice $\sigma(x)=-(1+\tanh(x))/2$.  The optimization of the
data collapse was performed using the simulated annealing
procedure. An example of the result is given in Fig.
\ref{fig:randbip-ipr-0.5-18-50-detail-scaling}, using the same data as
shown in Fig. \ref{fig:randbip-ipr-0.5-18-50-detail}. We can see that
the data collapse looks very good.}

\begin{figure}[t]
\includegraphics[scale=0.85]{%
\slaninafigdir/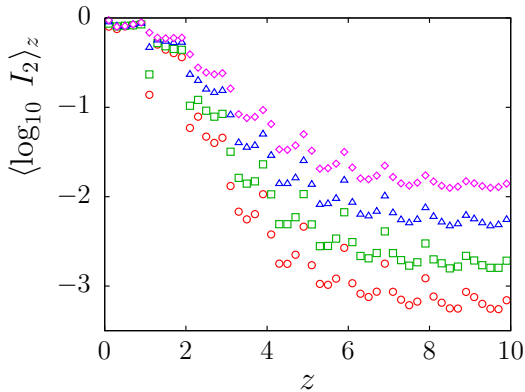}
\caption{Detail of the averaged logarithm of the inverse participation ratio for the graph
with $\alpha=1/2$, $M/N=18$ and $L/N=10$ and sizes $N=30000$ ({\Large $\circ$}),
  $N=10000$ ($\Box$), $N=3000$ ($\triangle$), and $N=1000$ ({\Large
  $\diamond$}). 
}
\label{fig:randbip-ipr-0.5-18-10-detail}
\end{figure}

{From thus obtained estimates
$z_c(N)$ the mobility edge should be extrapolated in the limit
$N\to\infty$. We found that the best fit of the size dependence
provides the formula $z_c(N)=z_\mathrm{mob}+aN^{-1/4}$ with some
constant $a$. An example of the fit is shown in the inset of Fig.
\ref{fig:randbip-ipr-0.5-18-50-detail-scaling}. This way we obtain the
mobility edge for all graphs studied. However, for a range of
parameters the estimated mobility edge falls below zero, which means
that localization is not observed at all. This happens typically for
small values of $L/N$, i. e. if the graph is very sparse.
An example of such
situation is shown in Fig. \ref{fig:randbip-ipr-0.5-18-10-detail},
where $L/N=10$ and the mobility edge determined by the above procedure
is negative, so we conclude that localization is absent. However, the
presence of delta-functions at integer values of $z$ makes the
analysis delicate and a more sophisticated procedure would be perhaps
desirable. }

\begin{figure}[t]
\includegraphics[scale=0.85]{%
\slaninafigdir/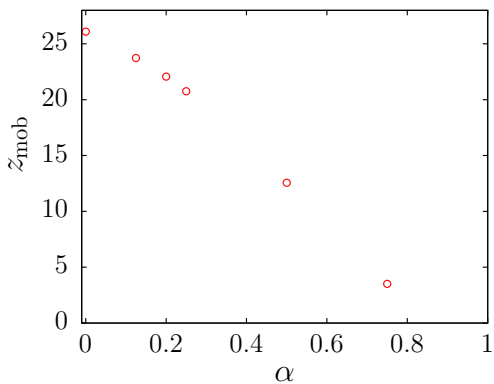}
\caption{Dependence of the estimated position of the mobility edge on the parameter
  $\alpha$, for graphs with parameters $M/N=18$ and $L/N=50$.
}
\label{fig:randbip-mobility-edge-vs-alpha}
\end{figure}

\begin{figure}[t]
\includegraphics[scale=0.85]{%
\slaninafigdir/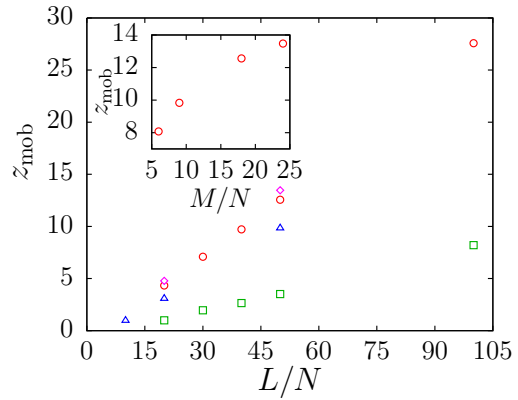}
\caption{Dependence of the estimated position of the mobility edge on the parameter
  $L/N$ for $\alpha=1/2$ and $M/N=18$ ({\Large $\circ$}), $\alpha=1/2$ and $M/N=24$
  ({\Large $\diamond$}), $\alpha=1/2$, and $M/N=9$ ($\triangle$), $\alpha=3/4$ and
  $M/N=18$ ($\Box$). In the inset, the dependence on $M/N$ for
  $\alpha=1/2$ and $L/N=50$.
}
\label{fig:randbip-mobility-edge-vs-k}
\end{figure}

The dependence of { the critical value
$z_\mathrm{mob}$} on graph parameters is shown in
Figs. \ref{fig:randbip-mobility-edge-vs-alpha} and
\ref{fig:randbip-mobility-edge-vs-k}. 
First we can see that when the exponent $\alpha$ increases, the
{  value of $z_\mathrm{mob}$}  moves toward zero,
until it disappears before $\alpha$ reaches the value $\alpha=1$.

Dependence on the parameter $L/N$  is shown in
Fig. \ref{fig:randbip-mobility-edge-vs-k}. We can observe more or less
linear dependence on $L/N$, and vanishing of $z_\mathrm{mob}$ at
certain value of this ratio, which is about $L/N\simeq 10$ for
$\alpha=1/2$, $M/N=18$ and $L/N\simeq 15$ for $\alpha=3/4$, $M/N=18$. The
dependence on the ratio $M/N$, as testified in the inset of
Fig.  \ref{fig:randbip-mobility-edge-vs-k}, shows that the position of
the mobility edge $z_\mathrm{mob}$ diminishes when the ratio
$M/N$ approaches one.

In fact, it can be clearly observed  that at the jamming threshold, $M/N=1$, 
the mobility edge  disappears, as it is demonstrated in the
inset of Fig. \ref{fig:randbip-dos-0.5-1-50}. This is consistent with
the view of jamming threshold as a critical point. When we approach
the critical point, the characteristic length scale diverges and as
soon as it surpasses the localization length, localization is gone. At
the same time we must be aware of the fact that jamming in granular matter occurs
in three- or two-dimensional Euclidean space, while the random graph
model of this article is not embedded in any Euclidean dimension. So,
the qualitative considerations based on length scales surely cannot
capture the full depth of the localization-versus-jamming problem.

\begin{figure}[t]
\includegraphics[scale=0.85]{%
\slaninafigdir/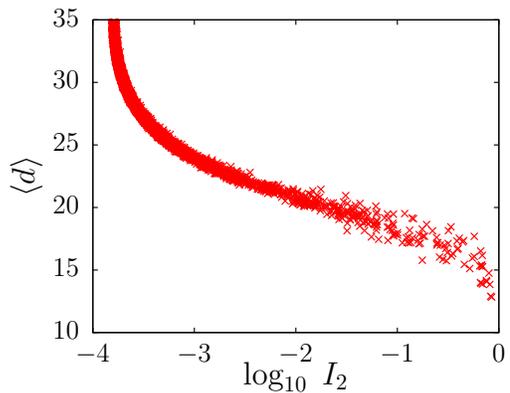}
\caption{This graph shows correlation between average degree $\langle
  d\rangle$ and inverse participation ratio. Each point corresponds to
  a single eigenvector in the lower part of the spectrum, $z<30$. The
  parameters of the graph are $\alpha=0.5$, $N=30000$, 
  $M/N=18$, $L/N=50$. 
}
\label{fig:randbip-degree-vs-ipr-0.5-18-50}
\end{figure}

{A question which naturally occurs is how are the localized states
related to other structural properties of the graph. Within the set of
graphs investigated in this work we observed only quite strong
correlation with the degree of nodes on which localization occurs. If
$d_i$ is the degree of node $i$, we can average with respect to
normalized eigenvector $e_{iz}$ corresponding to eigenvalue $z$ as $\langle
d\rangle = \sum_id_i e_{iz}^2$. We can see typical behavior in the
lower part of the spectrum in
Fig. \ref{fig:randbip-degree-vs-ipr-0.5-18-50}. Clearly, the localized
modes are centered at nodes with small degrees. We found that this is
generic in the graphs studied here.}

\section{Multifractality}

\subsection{{Motivation}}

{
It is now well established that the eigenvectors at the localization
transition exhibit multifractality (for review, see
e. g. \cite{eve_mir_08}). This peculiarity makes the localization
transition more complex than usual critical phenomena in absence of
quenched disorder. Recently there were studies hinting at multifractal
statistics of eigenvectors also off criticality \cite{mon_gar_11}. 
In the context of many-body localization it was found that not only
the critical, but also the extended states exhibit multifractality
\cite{del_alt_kra_sca_14}. It is argued that this has decisive role in
non-ergodicity of extended states. This topic is under hot debate
currently \cite{tik_mir_skv_16,tik_mir_16,alt_iof_kra_16}. It seems that unusual
multifractal behavior is related to the fact that these studies are
carried on graphs containing randomness (Bethe lattices, random
regular graphs, etc.) rather than in Euclidean space of dimensionality
at most three. For example, the many-body localization occurs in Fock
space, which has complex graph topology, usually approximated by
locally tree-like graphs \cite{del_alt_kra_sca_14}. Therefore, we find
natural to ask what are the multifractal properties of eigenvectors 
on  random graphs of the specific type investigated here. 
}

\subsection{Definitions}

We shall use the notion of multifractality in a slightly modified
sense, which we consider more appropriate for the problem at hand. To
keep our language clear, let us introduce our definitions together with
a few trivial examples.

The key quantities will be the moments of the eigenvectors 
$I_q(z)=\sum_{i=1}^N e_{iz}^{2q}$, where $q$ can assume any
positive as well as negative value. For $q=2$ we recover the usual
inverse participation ratio and for $q=1$ we have  $I_1(z)=1$ for all
eigenvalues $z$ due
to normalization. in order to compare the values at different graph
sizes, we should average over eigenvalues lying inside a narrow
interval $(z_-,z_+)$ centered at a fixed value $z=(z_-+z_+)/2$, exactly
as it was when investigating the inverse participation ratio. So, 
$\langle I_q\rangle_z=\sum_{z'\in(z_-,z_+)}I_q(z')/N_z$,
where  $N_z$ is the number of eigenvalues inside the interval $(z_-,z_+)$.

When studying the multifractal properties of the eigenvectors, 
we suppose that the averages scale with the graph size as $\langle
I_q\rangle_z\sim N^{-\zeta(q)}$, when $N\to\infty$. The function
$\zeta(q)$ embodies the information on the multifractal character of
the eigenvectors whose eigenvalues lie close to the point $z$. Let us
see what the function $\zeta(q)$ looks like in a benchmark case, which
is the Gaussian orthogonal ensemble. The distribution of eigenvector
elements is Gaussian \cite{bro_flo_fre_mel_pan_won_81,la_ci_bou_po_99}
 (also called Porter-Thomas distribution in this
context), which results in the following dependence on $N$
\cite{bro_flo_fre_mel_pan_won_81,eva_eco_90}. For  $q>-1/2$ we have
$\langle I_q\rangle_z\simeq
N^{1-q}\;\frac{\Gamma(2q+1)}{2^q\,\Gamma(q+1)}$ 
and for $q<-1/2$ we have  $\langle I_q\rangle_z\sim
N^{-3q}$. Therefore, for GOE
\begin{equation}
\zeta(q)=\min(3q,q-1)\;.
\label{eq:zetagoe}
\end{equation}

Let us now look at the eigenvectors with eigenvalues close to a fixed
value $z$ from a different perspective.
We assume that the set of $N$ nodes can be divided into $G$ groups of
sizes $N_g$, $g=1,2,\dots,G$, according to the scaling of the
eigenvectors with the
graph size $N$.  We suppose that the elements of the eigenvectors
scale like $|e_{i}|\simeq a_g N^{-h_g/2}$ for all $i$ within the group $g$,
while the size of the group scales like $N_g\simeq b_g N^{d_g}$. 
The moments
of the eigenvector then behave as $\langle I_q\rangle_z\simeq \sum_{g=1}^G
a_gb_g\,N^{d_g-qh_g}$. For very large $N$ this sum is dominated by a
single term with the maximum exponent, hence $\langle I_q\rangle_z\sim
N^{-\zeta(q)}$, where
\begin{equation}
\zeta(q)=\min_g (qh_g-d_q)\;.
\label{eq:zetabyminim}
\end{equation}

{Clearly, in a random graph the classification into such groups is
only schematic. But we can introduce the classification in somewhat
more formal way as follows. We fix a value $z$ and find an eigenvector
with eigenvalue closest to $z$. Then we order the elements of the
eigenvector in ascending order according to their modulus. Then, the
smallest has index $1$ while 
the largest has index $N$. 
 So we obtain a non-decreasing 
function. Then we numerically differentiate this function, so we
obtain an estimate of probability density for the modulus of
eigenvector elements.  We can plot together such functions for all graphs sizes $N$
studied, while the value of $z$ remains fixed. It can be better done
on logarithmic scale. Then we try to 
rescale the graphs so that they are shifted by 
$(h/2)\log N$ rightwards and $d\log N$ downwards. If all the graphs
have a common intersection, we can conclude that we identified one group
characterized by exponents $h$ and $d$. The coordinates of the
intersection correspond to the parameters $\log a$ and $\log b$. Of
course, we never do such procedure in reality, as we would need to
check infinite number of possible combinations of $d$ and $h$. This
description serves to the
sole purpose to put the classification into groups on a more solid grounds.}

\begin{figure}[t]
\includegraphics[scale=0.85]{%
\slaninafigdir/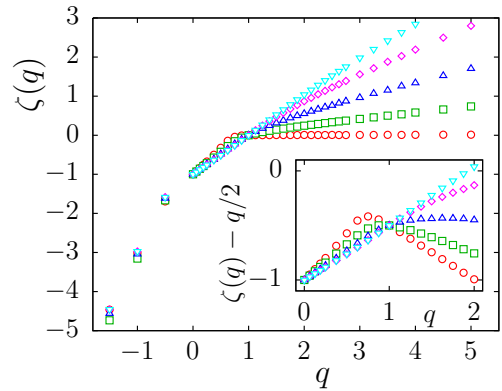}
\caption{Scaling exponents for the eigenvector moments $\langle
  I_q\rangle_z$. The averages are made over intervals of width $1$
  with midpoints at $z=8.5$ ({\Large $\circ$}), $z=11.5$ ($\Box$),
  $z=12.5$ ($\triangle$), $z=13.5$ ({\Large $\diamond$}), and $z=15.5$
  ({\large $\triangledown$}). In the inset, detail of the same
  data. The term $q/2$ was subtracted just for better visibility of
  the nonlinear dependence on $q$.
}
\label{fig:randbip-multif-exponents-0.5-18-50}
\end{figure}

We shall call the set of points $\{(h_g,d_g)|g=1,2,\ldots,G\}$ a multifractal
spectrum of the eigenvectors. We can write it as a function $d(h)$,
which can contain isolated points or  continuous part, or
both.
 Of course, it depends on the value $z$
around which the eigenvalues are taken.  To see the point, let us
consider eigenvectors of a complete graph. One of them (the ground
state) is totally delocalized, i. e. $e_i=N^{-1/2}$ and all the others
(excited states) are localized, one of them being
$e_1=\sqrt{(N-1)/N}$, $e_i=-1/\sqrt{(N-1)N}$, $i>1$. So, the
multifractal spectrum of the delocalized state consists of a single
point $(1,1)$, while the localized states have a pair of points
$\{(0,0),(2,1)\}$. In fact, the presence of the point $(0,0)$ is a
fingerprint of localization, as it implies that whole weight of the
eigenvector is carried by a set of sites which remains finite in the
thermodynamic limit. 

We can easily see by inspection that the multifractal spectrum
which results in the GOE exponents (\ref{eq:zetagoe}) is also composed
of just two points, $\{(1,1),(3,0)\}$. This is another example of a
trivial multifractality. To have a non-trivial multifractal spectrum,
or multifractality in proper sense, we need a continuous section in
the function $d(h)$. We shall see later that it occurs close to the mobility edge.

\begin{figure}[t]
\includegraphics[scale=0.85]{%
\slaninafigdir/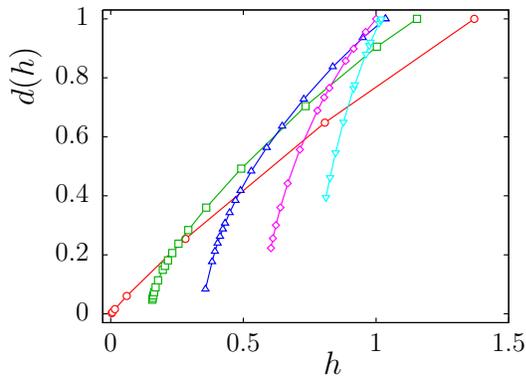}
\caption{Multifractal spectrum obtained numerically from the data
  shown in Fig. \ref{fig:randbip-multif-exponents-0.5-18-50}, with the
  same meaning of the symbols.
}
\label{fig:randbip-multif-spectrum-0.5-18-50}
\end{figure}

\subsection{Numerical results}

In our numerical studies, we
shall compute the multifractal spectrum from the calculated exponents $\zeta(q)$
by { a procedure which we call, for the sake of brevity, numerical
inversion of the equation (\ref{eq:zetabyminim}). The procedure goes
as follows. We are looking for the function $d(h)$, which might be
composed from a set of discrete points as well as continuous
part(s). First we should guess the interval $I$ into which all values
of $h$ should fall. As a first proxy for the function $d(h)$ we
calculate, for each $h\in I$, the location and value of the minimum 
\begin{equation}
v(h)=\min_q (qh-\zeta(q))=q_\mathrm{m}(h)h-\zeta(q_\mathrm{m}(h))\;.
\end{equation}
It would be misleading to identify the function $v(h)$ with $d(h)$, as
this would miss the fact that $d(h)$ contains isolated
points. To cure this problem, we identify $d(h)=v(h)$ for each $h$
except such points where $q_\mathrm{m}(h)$ is locally constant,
i. e. its first derivative with respect to $h$ exists and is zero. At
such points the function $d(h)$ is undefined. Of course, numerically
we  discretize the interval $I$ and check if
$q_\mathrm{m}(h)$ is constant by comparing its value at neighboring
points. 
}

{
Using this procedure we are respecting the fact that  if the function $\zeta(q)$
contains a linear piece, such piece in its entirety corresponds to a single isolated
point in the multifractal spectrum. 
}

We calculated the exponents $\zeta(q)$ for $z$ above and slightly
below the mobility edge. The density of states deep below the mobility
edge is too small to provide reasonable statistical error for
extracting the exponents from the averages $\langle I_q\rangle_z$. We show the 
results in Fig. \ref{fig:randbip-multif-exponents-0.5-18-50}. We
observe that $\zeta(q)=0$ for $q>1$ below the mobility edge, as it
should be in the localized state. Sufficiently far above the mobility
edge we observe $\zeta(q)=q-1$ for all positive $q$, i. e. the GOE
result. However, when we proceed from the mobility edge up, we 
depart from the localized behavior and approach the GOE limit rather
slowly, indicating a relatively wide interval of eigenvalues with
non-trivial behavior. The strong non-linearity of the function
$\zeta(q)$ is stressed by plotting the detail in the inset of Fig.
\ref{fig:randbip-multif-exponents-0.5-18-50}.  Moreover, we should
note that for negative exponents, more precisely for $q\le -1/2$, the
exponents obey the dependence $\zeta(q)=3q$ for all $z$, i. e. both in
the localized and delocalized phase. This is in fact the same behavior
as found also in GOE. So, this segment of the function $\zeta(q)$ is
very robust and is not 
influenced by localization at all.

We further analyzed the exponents by numerically inverting the
formula  (\ref{eq:zetabyminim}). The results are shown in
Fig. \ref{fig:randbip-multif-spectrum-0.5-18-50}. We can clearly see
that the function $d(h)$ reaches maximum value $d(h)=1$, as it
should. In the localized phase it contains also the point $(0,0)$, which
is fully consistent with the considerations above. The most
interesting part is the broad continuous 
section of he function $d(h)$, observed in the delocalized phase (note
that the point $(0,0)$ is not included!), not too far from the
mobility edge but certainly above it. This is the non-trivial part of
the multifractal spectrum. The width of the continuous part shrinks as
we go farther from the mobility edge, until it collapses onto the
single point $(1,1)$, characteristic of GOE. We could
speculate that the non-trivial multifractal spectrum in the
delocalized phase describes analogous phenomenon as found in
Ref. \cite{del_alt_kra_sca_14}. We should also note that the spectrum contains
also the isolated point $(3,0)$, although it is not shown in the figure. This
point originates from the behavior 
$\zeta(q)=3q$ for $q<-1/2$ which is shared with GOE always.

\begin{figure}[t]
\includegraphics[scale=0.85]{%
\slaninafigdir/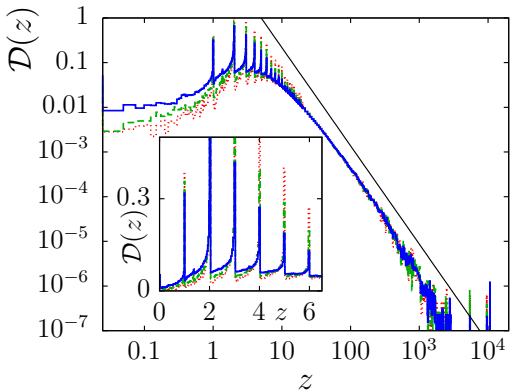}
\caption{Density of eigenvalues obtained in the analysis of the
  empirical Amazon network. The size of the subset is $N=30000$ (solid
  line), $N=16000$ (dashed line), and $N=10000$ (dotted line). The
  straight line is the power $\propto z^{-2.2}$. In the inset, detail
  of the same data.
}
\label{fig:randbip-amazon-dos}
\end{figure}

\begin{figure}[t]
\includegraphics[scale=0.85]{%
\slaninafigdir/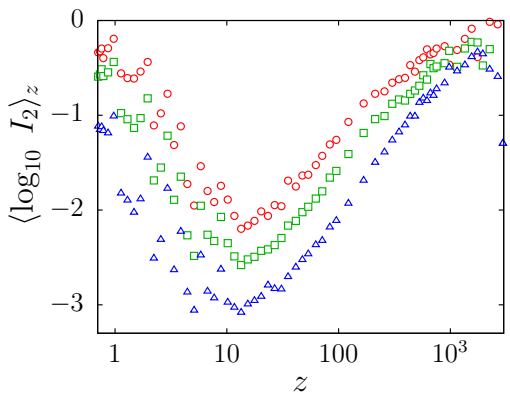}
\caption{Averaged logarithm of the inverse participation \\ratio, for
  the empirical Amazon network.  The size of the subset is $N=30000$
  ($\triangle$), $N=16000$ ($\Box$), and $N=10000$ ({\Large $\circ$}).
}
\label{fig:randbip-amazon-ipr}
\end{figure}

\section{Empirical study}

Now we turn to the analysis of an example of a bipartite scale-free
graph taken from reality. We utilize the same data collection as
already used in our previous work \cite{sla_kon_10}, so we only
briefly describe them now. The data were collected in the year 2005,
by automatic download from the server amazon.com. We were interested in
the network connecting reviewers and items offered on Amazon. This is
a bipartite graph, the set A being the collection of reviewers, the
set B is the set of items, and the edges represent the reviews
written. First we downloaded
the list of all 1,714,512 reviewers present in the system at that
time (now the list is hidden for its most part). The list was ordered
by Amazon itself by relevance, which meant more or less by number of
reviews written. Then, we downloaded
systematically all reviews written by the first $10^5$ reviewers in this
list. We found occasional duplicities in the data, and after cleaning
them we obtained a graph composed of 99,622 reviewers and 645,056
items. These two sets were connected by 2,036,091 reviews.

{In our study  \cite{sla_kon_10} we looked at eigenvectors with largest
IPR. We found that they identify small clusters within the network with
clear semantic content, e. g. groups of publications about certain
globally influential politician, another groups centered at popular
musical bands etc. It would be tempting to make such an
analysis automatically and rely on such computer-extracted
data. However, the very question of reliability of these data is
highly non-trivial. The first and essential question is whether the
localized states are just casual products of the randomness of the
underlying graph, or they are specific to this single empirical
network. We are trying to contribute to solving this question first by
comparing the Amazon network to a model random graph (which was done
in the preceding sections). Second, we proceed by taking the graph
representing the 
Amazon network as an input and trying to identify what is generic to
this type of graph (which is what we are about to do in this section).}

{Therefore, in order to investigate systematic properties of this graph, rather
than properties of this single empirical sample, we need an ensemble
of random graphs with structure as close as possible to the given
empirical sample. We prepare such random graphs by simply randomly selecting
subgraphs of the   empirical graph.} 
So, smaller
subgraphs were created by randomly choosing a set of $N$ reviewers and
leaving only items connected to them. In our studies we used three
sizes $N=10000$, $N=16000$, and $N=30000$. For each size we created 20
independent random realizations of the subset and averaged the density of
eigenvalues and inverse participation ratio in the same way as it was
done for the artificial graphs examined in the previous sections. Contrary
to the artificial case, here we cannot choose neither the size $M$ nor
the number of edges $L$ independently. These numbers also have some,
although small, sample-to-sample fluctuation. On average, we found that
$M/N\simeq 10$, for $N=30000$, $M/N\simeq 11.5$ for $N=16000$, and
$M/N\simeq 13$ for $N=10000$. The ratio $L/N$ is, for all three sizes,
$L/N\simeq 20$.  As we have already shown in \cite{sla_kon_10}, the
degree distribution is power-law on both the A and B sides, $P_{A,B}^>(k)\sim
k^{-\gamma_{A,B}}$, but the
exponents slightly differ: we found $\gamma_A\simeq 1.2$ and
$\gamma_B\simeq 1.35$. The density of eigenvalues is shown in Fig.   
\ref{fig:randbip-amazon-dos}. We can observe the peaks at integer
values of $z$ and the power-law tail. Comparing the spectra at
increasing graph sizes we observe that the convergence is quite slow at
small $z$, which is probably due to the fact that the ratio $M/N$ is
not quite the same for all sizes, but decreases slightly when $N$
increases. However, the tail seems not to be affected, as it relies
only on the power-law distribution of degrees.
According to the general theory
\cite{nagao_13}, only the smaller of the two exponents $\gamma_A$ and
$\gamma_B$ is relevant for
the tail of the eigenvalue density, so we expect that the exponent
will be $\tau\simeq 1.2$. We can see in
Fig. \ref{fig:randbip-amazon-dos} that this is very well 
confirmed by the data.    

\begin{figure}[t]
\includegraphics[scale=0.85]{%
\slaninafigdir/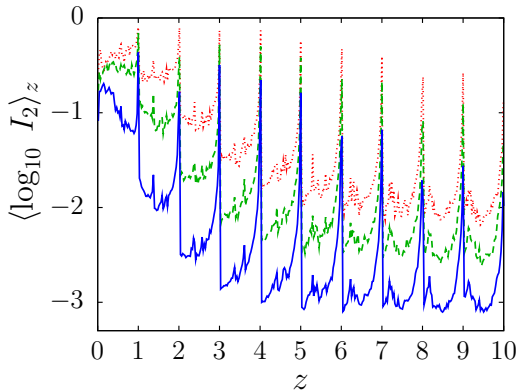}
\caption{Detail of the data shown in
  Fig. \ref{fig:randbip-amazon-ipr}.  The size of the subset is $N=30000$ (solid
  line), $N=16000$ (dashed line), and $N=10000$ (dotted line). 
}
\label{fig:randbip-amazon-ipr-detail1}
\end{figure}

In order to see if localized states occur in the spectrum, we plot
in Fig. \ref{fig:randbip-amazon-ipr-detail1} the inverse participation
ratio. In fact, no localization is observed in either low or high range of
eigenvalues. {At the upper tail, we observe qualitatively the same
behavior as in the model graphs investigated in previous
sections. There seems to be a crossover value $z_\mathrm{cross}$ and
all states within the tail but with $z<z_\mathrm{cross}$ are extended.
At the same time, $z_\mathrm{cross}$ seems to go to infinity with
growing graph size. However, the data are too noisy to see this effect
clearly. Neither the scaling like in (\ref{eq:scalingphi}) can be well
observed with the present data. So, the absence of localization in the
upper tail remains on the level of hypothesis, even more so than in the
case of model graphs. 
}

{
Again, at low eigenvalues the situation is more clear. It is obvious
that there is no region of localized states. }
The seemingly noisy behavior at small $z$ is in fact due
to complex features, which are repeated at all sizes, just shifted
downwards for increased $N$. This is clearly observable in the detail
shown in Fig. \ref{fig:randbip-amazon-ipr-detail1}. This figure also
confirms that the localized phase is not present here. When we look at
the behavior of the mobility edge in
Fig. \ref{fig:randbip-mobility-edge-vs-alpha}, we indeed expect
vanishing of localization for the parameters relevant for the
empirical data, which is, approximately, $M/N=10$, $L/N=20$,
$\alpha=3/4$. We can finally conclude that both eigenvalue density and
localization behavior of the empirical Amazon network is well reproduced
by the model of Goh et al. \cite{goh_kah_kim_01a} adapted for
bipartite graphs in \cite{nagao_13}.

\section{Conclusions}

We investigated localization in bipartite graphs with power-law degree
distribution. Localization is purely due to topological disorder. The
main quantity to characterize localization was the inverse 
participation ratio, more precisely its dependence on the graph
size.   In
comparison with the localization on Erd\H{o}s-R\'enyi graphs,
investigated by us earlier \cite{slanina_12b}, there are several
peculiarities. {First,  states at the upper tail of the
spectrum behave differently, which is demonstrated both in power-law
density of eigenvalues, and in localization properties. Based on our
data, we conjecture that 
all the states which are in the upper tail but with eigenvalue below certain
crossover value are extended; and, at the same time, the crossover value
goes to infinity when the graph size increases. Our current data seem
to support this hypothesis, but it would need better data to qualify
it as proved. }

At the lower edge of the spectrum, i. e. at eigenvalues close to zero,
the region of localized states remains intact in a generic
situation. However, the position of the mobility edge  $z_\mathrm{mob}$ depends
sensitively on the parameters of the model.  At certain values of the
parameters the critical value $z_\mathrm{mob}$  even drops to zero,
which means that 
localization disappears. The general trend is that  $z_\mathrm{mob}$ 
decreases with decreasing power in the degree distribution, with
decreasing density of edges in the graph and also drops to zero when
the sizes of the A and B sets (the two sides of the bipartite graph)
becomes equal. The latter case is especially important, because it
can be interpreted as jamming threshold in a man-field version of the
sphere packing problem. This
means that one critical point, the localization transition, is in
conflict with another critical point, the jamming transition. We
conjecture that this is due to competition between length scales
characteristic for the two transitions. When one length scale
prevails, the other critical point is concealed. There remains an
important open question of how the disappearance of low-lying
localized states will be reflected in heat conductance and sound
transmission near and at the jamming threshold. To proceed in solving
 this question 
it would be necessary to adapt the model so that it is embedded in a
three- or two-dimensional Euclidean space. 

{In the context of many-body localization \cite{bas_ale_alt_06} it is
common to study 
level-spacing statistics as an indicator for localization instead of
IPR  \cite{oga_hus_07}. However, we found that this method is not very
useful here, 
mainly due to very low density of states in the tail, where the
localization occurs. So, we do not use this method here (see
Appendix A for details).}

We also analyzed multifractal properties of the eigenvectors. In some
sense it is just deepening of the analysis based on the IPR. It
includes the estimate of the position of the mobility edge as a
by-product, as we identified the localized phase by the presence of
the point $(0,0)$ in the multifractal spectrum. However, the finding
we consider interesting is that the multifractal spectrum is
non-trivial next to but in relatively broad range
above the mobility edge. This may indicate that the eigenvectors are multifractal in
the delocalized phase, as was found on random regular graphs in
Ref. \cite{del_alt_kra_sca_14} or with a different approach in
Ref. \cite{bir_tei_tar_12}, although these conclusions were questioned
in \cite{tik_mir_skv_16}. In our view this phenomenon is connected
with the topology of the random graph, more precisely with the
distribution of loop lengths in the graph. Indeed, the graph is locally
tree-like, typical loops having length $\ln N$. For computing spectra
this is sufficient, but localization (or rather delocalization) is
essentially non-local phenomenon and loops of length $\ln N$ cannot be
considered large close to the critical region, where a typical length
scale originating from localization competes with the typical length
scale $\ln N$ originating from the topology. In finite-dimensionality
lattices, 
distribution of loop 
lengths does not depend on system size, while in our random graph it
does. Therefore, the scaling of the moments $I_q$ with graph size may
be non-trivial in the delocalized phase. However, if this hypothesis
is true, the range of observed multifractal states above mobility edge would
ultimately shrink when system size grows, although for numerically
accessible sizes the ultimate shrinking to a point may never be
observable.  

As a complement to the study of artificial bipartite graphs, we
analyzed also one empirical bipartite graph, namely the network of
reviewers and items on the amazon.com server. In our previous study
\cite{sla_kon_10} we observed the scale-free nature of this graph and
by extracting the most localized eigenvectors we found small
communities with sensible semantic information. Here we wanted to
check if the model of Goh et al. is useful also in describing the
spectra of the graphs and properties of its eigenvectors. We found
that indeed, the spectral and localization properties found in the
empirical network are reproduced well within the Goh et
al. model. This makes of it a useful benchmark in spectral studies of
empirical networks, at least the electronic commerce networks
investigated in \cite{sla_kon_10}.
{Most important conclusion is that
in the empirical graph there should be no localization due to purely
random geometry of the network. Therefore, the localized states found
in  \cite{sla_kon_10} are true outliers who bear specific information
on this single instance of the empirical network. This is what we
took as an assumption in  \cite{sla_kon_10} and now we believe it is
more firmly supported by the analysis of this work.}

\begin{acknowledgments}
I wish to thank  K. Neto\v{c}n\'y for fruitful discussions.

\end{acknowledgments}
\appendix

\section{Eigenvalue spacings}

One of the key differences between localized and delocalized portion
of the spectrum consists in the statistics of eigenvalue spacings. In
GOE, it is very well approximated by the Wigner surmise
$P_{\Delta\,\mathrm{GOE}}(x)=\frac{\pi}{2}\, xe^{-\pi\,x^2/4}$, where $x$ is the
eigenvalue spacing normalized to its average value. If the eigenvalues
were placed randomly according to a Poisson process, the distribution
of normalized spacings would follow the exponential
$P_{\Delta\,\mathrm{Poisson}}(x)=e^{-x}$. General expectation is that
GOE result should hold in the delocalized regime, while localized
states should correspond to Poisson level spacing distribution. Let us
see now how this expectation is fulfilled in the case of our graphs.

\begin{figure}[h]
\includegraphics[scale=0.85]{%
\slaninafigdir/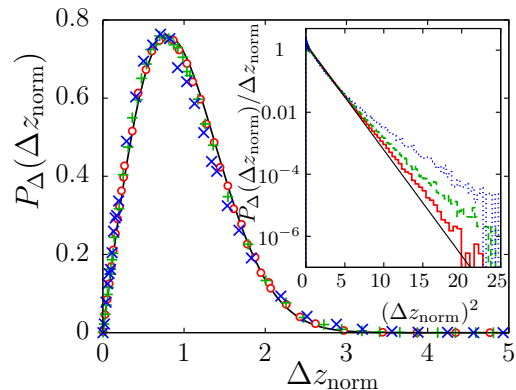}
\caption{Distribution of eigenvalue spacings, for eigenvalues
  within the interval $(20,22)$ ({\large $\circ$}, solid line in
  inset), $(101,111)$ ($+$, dashed
  line in inset), and $(200,210)$ ($\times$, dotted line in inset). In
  the inset, the same data are plotted in rescaled form, to show the
  behavior in the tail. The parameters of the graph
  are $\alpha=1/2$, $M/N=18$, $L/N=50$ and size $N=1000$. The full
  line in the main plot and the straight line in the inset are the
  Wigner surmise  $P_\Delta(\Delta z_\mathrm{norm}) =(\pi \Delta
   z_\mathrm{norm}/2)\exp(-\pi\,(\Delta
  z_\mathrm{norm})^2/4)$.} 
\label{fig:randbip-dist-hist-20-101-200-0.5-18-50}
\end{figure}
\begin{figure}[t]
\includegraphics[scale=0.85]{%
\slaninafigdir/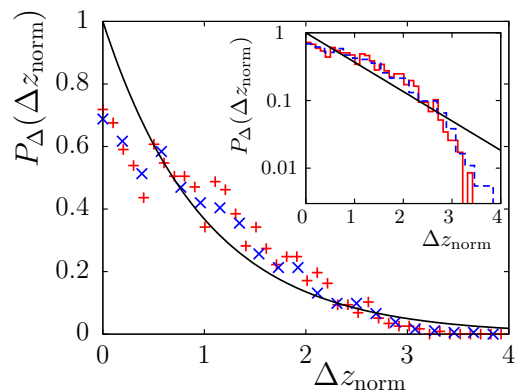}
\caption{Distribution of eigenvalue spacings, for eigenvalues
  within the interval $(8,9)$.  The parameters of the graph
  are $\alpha=0.5$, $M/N=18$, $L/N=50$ and size $N=1000$ ($+$, solid
  line in the inset), $N=3000$ ($\times$, dashed  line in the inset).
  In   the inset, the same data are plotted in logarithmic scale, to show
  the   behavior in the tail. The full line in the main plot and the
  straight line in the inset is the dependence $\exp(-\Delta
  z_\mathrm{norm})$. 
}
\label{fig:randbip-dist-hist-8-0.5-18-50}
\end{figure}
\begin{figure}[t]
\includegraphics[scale=0.85]{%
\slaninafigdir/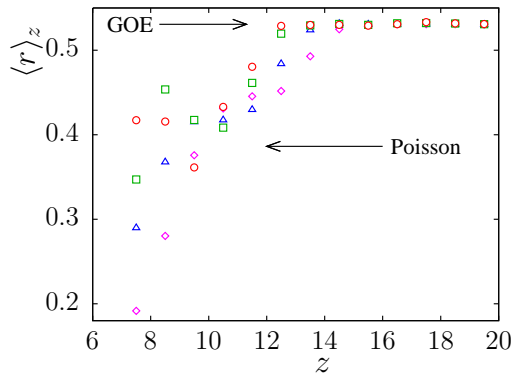}
\caption{Spacing variation parameter, averaged over eigenvalues within
  the interval of width $1$ centered at $z$.  The
  parameters of the 
  graph are $\alpha=1/2$, $M/N=18$, $L/N=50$ and sizes  $N=30000$ ({\Large $\circ$}),
  $N=10000$ ($\Box$), $N=3000$ ($\triangle$), and $N=1000$ ({\Large $\diamond$}).
The horizontal arrows indicate the values which correspond to Gaussian
orthogonal ensemble and to the Poisson placement of eigenvalues.}
\label{fig:randbip-dist-mom-r-8-0.5-18-50}
\end{figure}

We calculated the distribution of eigenvalue spacings within several
intervals across the spectrum. Taking interval $(z_-,z_+)$, we
normalized the spacing between adjacent eigenvalues $z_{i+1}$, $z_i$
as $\Delta z_\mathrm{norm}=\frac{N_z}{z_+-z_-}(z_{i+1}-z_i)$, where
$N_z$ is the number of eigenvalues in the interval $(z_-,z_+)$. Such
distributions can be directly compared with the GOE and/or Poisson
result. This is done in
Fig. \ref{fig:randbip-dist-hist-20-101-200-0.5-18-50} in the region
well above the 
mobility edge and in the tail of the spectrum, and in
Fig. \ref{fig:randbip-dist-hist-8-0.5-18-50}  in the
region slightly below the mobility edge. Deeper below the mobility
edge the analysis is hindered by very small density of eigenvalues. 

We can see that in the interval $(20,22)$, which lies above the
mobility edge in the region of high density of eigenvalues, level
spacings follow very well the GOE formula, confirming the status of
delocalized eigenvectors. Farther in the tail, in the interval
$(101,111)$, and even more in $(200,210)$, there are deviations from
the Wigner surmise; large spacings are more probable than what GOE
predicts. But the deviations are relatively small and do not harm the
overall picture that all the delocalized regime is well characterized by
GOE level spacings.

Below the mobility edge the conclusions are much less clear. The
interval $(8,9)$ investigated here lies next to the mobility edge,
while the localized states lying farther are too rare to obtain
reasonable statistics of the level spacings. The distribution shown in
Fig.  \ref{fig:randbip-dist-hist-8-0.5-18-50} is certainly much closer
to Poisson than to GOE, confirming the prediction that in localized
regime GOE breaks down. However, we still cannot claim that the
Poisson level spacing would make a good fit to the measured data. We
assume that the difference from Poisson is due to closeness of the
mobility edge. Farther away the spacing distribution is expected to
correspond to the Poisson case much better.

\begin{figure}[t]
\includegraphics[scale=0.85]{%
\slaninafigdir/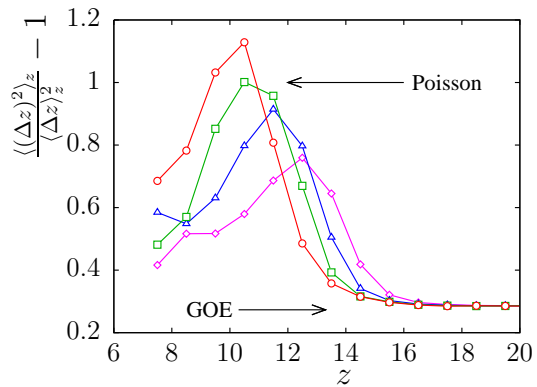}
\caption{Relative variance of the eigenvalue spacing distribution within
  the interval of width $1$ centered at $z$. 
 The
  parameters of the 
  graph are $\alpha=1/2$, $M/N=18$, $L/N=50$ and sizes  $N=30000$ ({\Large $\circ$}),
  $N=10000$ ($\Box$), $N=3000$ ($\triangle$), and $N=1000$ ({\Large $\diamond$}).
The horizontal arrows indicate the values which correspond to Gaussian
orthogonal ensemble and to the Poisson placement of eigenvalues.
}
\label{fig:randbip-dist-mom-fluc-8-0.5-18-50}
\end{figure}

Besides the full spacing distribution, there are aggregate parameters
characterizing the spacing distribution in terms of a single number.
The first one we use here is the ratio of two consecutive spacings
$\Delta z_i=z_i-z_{i-1}$ and $\Delta z_{i+1}=z_{i+1}-z_i$
defined as \cite{oga_hus_07}
\begin{equation}
r=\frac{\min(\Delta z_i,\Delta z_{i+1})}{\max(\Delta z_i,\Delta z_{i+1})}
\end{equation}
and used frequently in the context of many-body localization
\cite{bas_ale_alt_06}. This quantity was averaged over an interval
of eigenvalues cantered at $z$, as it was done with inverse
participation ratio. The average $\langle r\rangle_z$ should reflect
the transition from GOE behavior, where $\langle r\rangle_z=0.529\ldots$,
to Poisson behavior, where  $\langle r\rangle_z=2\ln
2-1=0.386\ldots$ \cite{oga_hus_07}. We can see the results in
Fig. \ref{fig:randbip-dist-mom-r-8-0.5-18-50}. In the region far above
the mobility edge we indeed observe that $\langle r\rangle_z$ settles
at the GOE value. When we approach the mobility edge,   $\langle
r\rangle_z$ decreases, indicating the transition. However, the
behavior below the mobility edge is ambiguous. The density of states
is so small that the statistical fluctuations obscure the
trend. Essentially the data are compatible with the idea that $\langle
r\rangle_z$ approaches the Poisson value, but a firm statement cannot
be done on the basis of current data. Certainly our results in no
means prove that the spacing distribution in localized phase is
Poisson, although the results neither prove the contrary. 

Another single-valued indicator capable in principle to discern
between GOE and Poisson regimes is the relative variance of the spacing
distribution $\langle(\Delta z)^2\rangle_z/\langle\Delta z\rangle^2_z-1$. this
quantity should be equal $1$ for Poisson and $4/\pi-1=0.2732\ldots$
for GOE spacing. We can see the numerical results for our graphs in
Fig. \ref{fig:randbip-dist-mom-fluc-8-0.5-18-50}. The behavior
resembles that of the quantity $\langle r\rangle_z$. For $z$
sufficiently above the mobility edge, the relative variance is
precisely at the GOE value. When we approach the mobility edge, the
relative variance grows and at the mobility edge and below it
decreases again. However, instead of approaching the Poisson limit, it
is significantly smaller. This again casts some doubts at the hypothesis
that in the localized phase the spacing distribution is Poisson.
However, the discrepancy may also be due to finite size effects. As we
can see in Fig.  \ref{fig:randbip-dist-mom-fluc-8-0.5-18-50}, the
function does not seem to converge well in the range of graph sizes
studied. 

One should ask how to interpret these, rather negative, results. 

Usually it is expected that the statistics corresponds to the
Gaussian orthogonal ensemble in the delocalized regime, while it is
Poisson in the localized one. The  hypothesis on the delocalized state
was fully
confirmed in our studies, both by directly plotting the spacing
distributions and by using aggregate quantities, like adjacent spacing
ratio and relative variance. The deviations from GOE, if they exist at
all, are small and restricted to the region of extremely large
spacings, where finite size may play role. 

On the other hand, the results in the localized regime are much less
conclusive, mainly because the density of states is too low to obtain
reliable results deep within the localized region. Below, but close to,
the mobility edge, the spacing distribution is close to Poisson in the
sense that it is purely decreasing function, but decreases more
rapidly than an exponential, as would be expected for Poisson
case. Also the aggregate quantities behave in a similar way; their
value clearly departs from the GOE value when the mobility edge is
crossed, but remains rather far from the value expected by Poisson
spacing. We conclude that the analysis of eigenvalue spacing is not
quite distinctive method for analyzing localization in this
system. This is in contrast with the situation in e. g. random regular
graphs studied by us in \cite{slanina_12b}, where the Poisson
distribution in localized phase was very well visible. We attribute
this difference to the already mentioned scarcity of eigenvalues in
localized region, which is characteristic for topological disorder
(both in Erd\H{o}s-R\'enyi graphs studied in \cite{slanina_12b} and in
bipartite graphs studied here). On the contrary, the random regular
graphs are dominated by diagonal disorder (i. e. random potential) and
the topological disorder, which is also present, has negligible effect.
This has the effect that at in the localized phase there is still
quite large density of eigenvalues, or even all the eigenstates are
localized. This implies that reliable analysis of localization in
topological disorder is much harder that in diagonal disorder.

The problem may be also formulated in the following way. Clearly,
localization is an effect which cannot be in principle analyzed
without a finite-size analysis. Indeed, localization means that the
extent of the state does not increase when the extent of the whole
system increases. Change in the level statistics seemingly avoids the
necessity of both calculating IPR and making finite-size
analysis. However, level statistics is only a secondary indicator. It
is supposed to change abruptly when we are already working with
infinite system. For finite systems we do not have any clue of how the
spacing distribution should scale when the system size changes. The
source of the difference between Poisson and GOE spacing distribution
lies in the absence or presence of level repulsion in localized and
delocalized states, respectively. When
the density of eigenvalues is small, level repulsion can be hardly
effective, thus masking the Poisson-to-GOE transition. This sheds some
doubts on the use of aggregate quantities like the spacing ratio, but more
detailed methodological analysis would be necessary here.

\section{Eigenvector of $R$ follows from eigenvector of $C$}

Assumption is 
\begin{equation}
  Ce=\lambda^2e\;.
\end{equation}
Denote
\begin{equation}
  \bar{e}=
  \left(
  \begin{array}{c}
    e\\
    S^Te/\lambda
  \end{array}
  \right)\;.
\end{equation}
Therefore
\begin{equation}
  \begin{split}
  R\bar{e}=\left(
\begin{array}{cc}
0 & S\\
S^T & 0
\end{array}
\right)
\left(
\begin{array}{c}
  e\\
  S^Te/\lambda
\end{array}
\right)=
\left(
\begin{array}{c}
  SS^Te/\lambda\\
  S^Te
\end{array}
\right)=\\
\left(
\begin{array}{c}
  Ce/\lambda\\
  \lambda S^Te/\lambda
\end{array}
\right)=
\lambda\left(
\begin{array}{c}
  e\\
   S^Te/\lambda
\end{array}
\right)=\lambda\,\bar{e}\;.
\end{split}
\end{equation}

\begin{thebibliography}{99}
\bibitem{anderson_58}
P. W. Anderson,
Phys. Rev.
 {\bf 109},
 1492
 (1958).

\bibitem{abrahams_10}
E. Abrahams (Ed.),
{\it 50 Years of Anderson Localization}
 (World Scientific, Singapore, 2010).

\bibitem{lee_ram_85}
P. A. Lee and T. V. Ramakrishnan,
Rev. Mod. Phys.
 {\bf 57},
 287
 (1985).

\bibitem{kra_mcki_93}
B. Kramer and A. MacKinnon,
Rep. Prog. Phys.
 {\bf 56},
 1469
 (1993).

\bibitem{eve_mir_08}
F. Evers and A. D. Mirlin,
Rev. Mod. Phys.
 {\bf 80},
 1355
 (2008).

\bibitem{ku_sou_83}
H. Kunz and B. Souillard,
J. Physique Letters
 {\bf 44},
 L-411
 (1983).

\bibitem{stollmann_01}
P. Stollmann,
{\it Caught by disorder. Bound states in Random Media}
 (Birkh\"auser, Boston, 2001).

\bibitem{abr_and_lic_ram_79}
E. Abrahams, P. W. Anderson, D. C. Licciardello, and T. V. Ramakrishnan,
Phys. Rev. Lett.
 {\bf 42},
 673
 (1979).

\bibitem{vol_wol_80}
D. Vollhardt and P. W\"olfle,
Phys. Rev. B
 {\bf 22},
 4666
 (1980).

\bibitem{suslov_95}
I. M. Suslov,
Sov. Phys. JETP
 {\bf 81},
 925
 (1995).

\bibitem{jan_kol_05}
V. Jani\v{s} and J. Koloren\v{c},
Phys. Rev. B
 {\bf 71},
 033103
 (2005).

\bibitem{wegner_79a}
F. J. Wegner,
Phys. Rev. B
 {\bf 19},
 783
 (1979).

\bibitem{efetov_83}
K. B. Efetov,
Adv. Phys.
 {\bf 32},
 53
 (1983).

\bibitem{ab_an_tho_73}
R. Abou-Chacra, P. W. Anderson, and D. J. Thouless,
J. Phys. C: Solid State Phys.
 {\bf 6},
 1734
 (1973).

\bibitem{ab_tho_74}
R. Abou-Chacra and D. J. Thouless,
J. Phys. C: Solid State Phys.
 {\bf 7},
 65
 (1974).

\bibitem{log_wol_85}
D. E. Logan and P. G. Wolynes,
Phys. Rev. B
 {\bf 31},
 2437
 (1985).

\bibitem{ant_eco_77}
P. D. Antoniou and E. N. Economou,
Phys. Rev. B
 {\bf 16},
 3768
 (1977).

\bibitem{gir_jon_80}
S. M. Girvin and M. Jonson,
Phys. Rev. B
 {\bf 22},
 3583
 (1980).

\bibitem{efetov_90}
K. B. Efetov,
Physica A
 {\bf 167},
 119
 (1990).

\bibitem{mi_fyo_91}
A. D. Mirlin and Y. V. Fyodorov,
Nucl. Phys. B
 {\bf 366},
 507
 (1991).

\bibitem{markos_06}
P. Marko\v{s},
Acta Physica Slovaca
 {\bf 56},
 561
 (2006).

\bibitem{mon_gar_10}
C. Monthus and T. Garel,
Phys. Rev. B
 {\bf 81},
 224208
 (2010).

\bibitem{bollobas_85}
B. Bollob\'as,
{\it Random Graphs}
 (Academic Press, London, 1985).

\bibitem{ja_na_be_96a}
H. M. Jaeger, S. R. Nagel, and R. P. Behringer,
Rev. Mod. Phys.
 {\bf 68},
 1259
 (1996).

\bibitem{maj_beh_05}
T. S. Majmudar and R. P. Behringer,
Nature
 {\bf 435},
 1079
 (2005).

\bibitem{ji_ca_ve_99}
X. Jia, C. Caroli, and B. Velicky,
Phys. Rev. Lett.
 {\bf 82},
 1863
 (1999).

\bibitem{owe_dan_11}
E. T. Owens and K. E. Daniels,
EPL
 {\bf 94},
 54005
 (2011).

\bibitem{bas_owe_dan_por_12}
D. S. Bassett, E. T. Owens, K. E. Daniels, and M. A. Porter,
Phys. Rev. E
 {\bf 86},
 041306
 (2012).

\bibitem{zel_poh_71}
R. C. Zeller and R. O. Pohl,
Phys. Rev. B
 {\bf 4},
 2029
 (1971).

\bibitem{xu_vit_wya_liu_nag_09}
N. Xu, V. Vitelli, M. Wyart, A. J. Liu, and S. R. Nagel,
Phys. Rev. Lett.
 {\bf 102},
 038001
 (2009).

\bibitem{ca_wi_bou_cla_98b}
M. E. Cates, J. P. Wittmer, J.-P. Bouchaud, and P. Claudin,
Phys. Rev. Lett.
 {\bf 81},
 1841
 (1998).

\bibitem{liu_nag_10}
A. J. Liu and S. R. Nagel,
Annu. Rev. Condens. Matter Phys.
 {\bf 1},
 347
 (2010).

\bibitem{liu_nag_saa_wya_11}
A. J. Liu, S. R. Nagel, W. van Saarloos, and M. Wyart,
in: {\it Dynamical Heterogeneities in Glasses, Colloids, and Granular Media}, eds. L. Berthier, G. Biroli, J.-P. Bouchaud, L. Cipelletti, and W. van Saarloos
 (Oxford UP, Oxford, 2011).

\bibitem{tor_sti_10}
S. Torquato and F. H. Stillinger,
Rev. Mod. Phys.
 {\bf 82},
 2633
 (2010).

\bibitem{ast_wea_08}
T. Aste and D. Weaire,
{\it The Pursuit of Perfect Packing}
 (Taylor and Francis, Boca Raton, 2008).

\bibitem{con_slo_99}
J. H. Conway and N. J. A. Sloane,
{\it  Sphere Packings, Lattices and Groups}
 (Springer, Berlin, 1999).

\bibitem{parisi_14}
G. Parisi,
arXiv:1401.4413.

\bibitem{fra_par_16}
S. Franz and G. Parisi,
J. Phys. A: Math. Theor.
 {\bf 49},
 145001
 (2016).

\bibitem{fra_par_urb_zam_16}
S. Franz, G. Parisi, P. Urbani, and F. Zamponi,
Proc. Nat. Acad. Sci. USA
 {\bf 112},
 14539
 (2015).

\bibitem{nishimori_01}
H. Nishimori,
{\it Statistical Physics of Spin Glasses and Information Processing}
 (Clarendon Press, Oxford, 2001).

\bibitem{alb_bar_01}
R. Albert and A.-L. Barab\'asi,
Rev. Mod. Phys.
 {\bf 74},
 47
 (2002).

\bibitem{la_ci_bou_po_99}
L. Laloux, P. Cizeau, J.-P. Bouchaud, and M. Potters,
Phys. Rev. Lett.
 {\bf 83},
 1467
 (1999).

\bibitem{ple_gop_ros_am_sta_99}
V. Plerou, P. Gopikrishnan, B. Rosenow, L. A. N. Amaral, and H. E. Stanley,
Phys. Rev. Lett.
 {\bf 83},
 1471
 (1999).

\bibitem{ciz_bou_94}
P. Cizeau and J.-P. Bouchaud,
Phys. Rev. E
 {\bf 50},
 1810
 (1994).

\bibitem{bur_jur_now_04}
Z. Burda, J. Jurkiewicz, M. A. Nowak, G. Papp,  and I. Zahed,
Physica A
 {\bf 343},
 694
 (2004).

\bibitem{sla_kon_10}
F. Slanina and Z. Konop\'asek,
Adv. Compl. Syst.
 {\bf 13},
 699
 (2010).

\bibitem{slanina_12a}
F. Slanina,
Adv. Compl. Syst.
 {\bf 15},
 1250053
 (2012).

\bibitem{slanina_14}
F. Slanina,
Adv. Compl. Syst.
 {\bf 17},
 1450002
 (2014).

\bibitem{sad_kal_hav_ber_05}
M. Sade, T. Kalisky, S. Havlin, and R. Berkovits,
Phys. Rev. E
 {\bf 72},
 066123
 (2005).

\bibitem{zhu_yan_yin_li_08}
G. Zhu, H. Yang, C. Yin, and B. Li,
Phys. Rev. E
 {\bf 77},
 066113
 (2008).

\bibitem{jal_sol_vat_li_10}
S. Jalan, N. Solymosi, G. Vattay, and B. Li,
Phys. Rev. E
 {\bf 81},
 046118
 (2010).

\bibitem{gir_geo_she_09}
O. Giraud, B. Georgeot, and D. L. Shepelyansky,
Phys. Rev. E
 {\bf 80},
 026107
 (2009).

\bibitem{odor_14}
G. \'Odor,
Phys. Rev. E
 {\bf 90},
 032110
 (2014).

\bibitem{far_der_bar_vic_01}
I. J. Farkas, I. Der\'enyi, A.-L. Barab\'asi, and T. Vicsek,
Phys. Rev. E
 {\bf 64},
 026704
 (2001).

\bibitem{goh_kah_kim_01b}
K.-I. Goh, B. Kahng, and D. Kim,
Phys. Rev. E
 {\bf 64},
 051903
 (2001).

\bibitem{dor_gol_men_sam_03}
S. N. Dorogovtsev, A. V. Goltsev, J. F. F. Mendes, and A. N. Samukhin,
Phys. Rev. E
 {\bf 68},
 046109
 (2003).

\bibitem{rod_bra_88}
G. J. Rodgers and A. J. Bray,
Phys. Rev. B
 {\bf 37},
 3557
 (1988).

\bibitem{sem_cug_02}
G. Semerjian and L. F. Cugliandolo,
J. Phys. A: Math. Gen.
 {\bf 35},
 4837
 (2002).

\bibitem{rod_aus_kah_kim_05}
G. J. Rodgers, K. Austin, B. Kahng, and D. Kim,
J. Phys. A: Math. Gen.
 {\bf 38},
 9431
 (2005).

\bibitem{nag_rod_08}
T. Nagao and G. J. Rodgers,
J. Phys. A: Math. Theor.
 {\bf 41},
 265002
 (2008).

\bibitem{cav_gia_par_99}
A. Cavagna, I. Giardina, and G. Parisi,
Phys. Rev. Lett.
 {\bf 83},
 108
 (1999).

\bibitem{kuhn_08}
R. K\"uhn,
J. Phys. A: Math. Theor.
 {\bf 41},
 295002
 (2008).

\bibitem{slanina_11}
F. Slanina,
Phys. Rev. E
 {\bf 83},
 011118
 (2011).

\bibitem{rod_ded_90}
G. J. Rodgers and C. De Dominicis,
J. Phys. A: Math. Gen.
 {\bf 23},
 1567
 (1990).

\bibitem{fyo_mir_91}
Y. V. Fyodorov and A. D. Mirlin,
J. Phys. A: Math. Gen.
 {\bf 24},
 2219
 (1991).

\bibitem{fyo_mir_91a}
Y. V. Fyodorov and A. D. Mirlin,
Phys. Rev. Lett.
 {\bf 67},
 2049
 (1991).

\bibitem{bi_mo_99}
G. Biroli and R. Monasson,
J. Phys. A: Math. Gen.
 {\bf 32},
 L255
 (1999).

\bibitem{cil_gri_mar_par_ver_05}
S. Ciliberti, T. S. Grigera, V. Mart\'{\i}n-Mayor, G. Parisi, and P. Verrocchio,
Phys. Rev. B
 {\bf 71},
 153104
 (2005).

\bibitem{rod_cas_kuh_tak_08}
T. Rogers, I. P\'erez Castillo,  R. K\"uhn, and K. Takeda,
Phys. Rev. E
 {\bf 78},
 031116
 (2008).

\bibitem{met_ner_bol_10}
F. L. Metz, I. Neri, and D. Boll\'e,
Phys. Rev. E
 {\bf 82},
 031135
 (2010).

\bibitem{bir_sem_tar_10}
G. Biroli, G. Semerjian, and M. Tarzia,
Prog. Theor. Phys Suppl.
 {\bf 184},
 187
 (2010).

\bibitem{mon_gar_11}
C. Monthus and T. Garel,
J. Phys. A: Math. Theor.
 {\bf 44},
 145001
 (2011).

\bibitem{kuh_mou_11}
R. K\"uhn and J. van Mourik,
J. Phys. A: Math. Theor.
 {\bf 44},
 165205
 (2011).

\bibitem{kuhn_16}
R. K\"uhn,
Phys. Rev. E
 {\bf 93},
 042110
 (2016).

\bibitem{slanina_12b}
F. Slanina,
Eur. Phys. J. B
 {\bf 85},
 361
 (2012).

\bibitem{evangelou_83}
S. N. Evangelou,
Phys. Rev. B
 {\bf 27},
 1397
 (1983).

\bibitem{evangelou_92}
S. N. Evangelou,
J. Stat. Phys.
 {\bf 69},
 361
 (1992).

\bibitem{eva_eco_92}
S. N. Evangelou and E. N. Economou,
Phys. Rev. Lett.
 {\bf 68},
 361
 (1992).

\bibitem{bir_tei_tar_12}
G. Biroli, A. C. Ribeiro-Teixeira, and M. Tarzia,
arXiv:1211.7334.

\bibitem{tik_mir_skv_16}
K. S. Tikhonov, A. D. Mirlin, and M. A. Skvortsov,
Phys. Rev. B
 {\bf 94},
 220203
 (2016).

\bibitem{nag_tan_07}
T. Nagao and T. Tanaka,
J. Phys. A: Math. Theor.
 {\bf 40},
 4973
 (2007).

\bibitem{kho_rod_97}
A. Khorunzhy and G. J. Rodgers,
J. Math. Phys.
 {\bf 38},
 3300
 (1997).

\bibitem{nagao_13}
T. Nagao,
J. Phys. A: Math. Theor.
 {\bf 46},
 065003
 (2013).

\bibitem{goh_kah_kim_01a}
K.-I. Goh, B. Kahng, and D. Kim,
Phys. Rev. Lett.
 {\bf 87},
 278701
 (2001).

\bibitem{lee_goh_kah_kim_04}
D.-S. Lee, K.-I. Goh, B. Kahng, and D. Kim,
Nucl. Phys. B
 {\bf 696},
 351
 (2004).

\bibitem{fle_sok_13}
F. Flegel and I. M. Sokolov,
Phys. Rev. E
 {\bf 87},
 022806
 (2013).

\bibitem{bau_gol_01}
M. Bauer and O. Golinelli,
J. Stat. Phys.
 {\bf 103},
 301
 (2001).

\bibitem{golinelli_03}
O. Golinelli,
cond-mat/0301437.

\bibitem{mar_pas_67}
V. A. Mar\v{c}enko and L. A. Pastur,
Math. USSR--Sbornik
 {\bf 1},
 457
 (1967).

\bibitem{her_sil_liu_nag_03}
C. S. O'Hern, L. E. Silbert, A. J. Liu, and S. R. Nagel,
Phys. Rev. E
 {\bf 68},
 011306
 (2003).

\bibitem{sil_liu_nag_05}
L. E. Silbert, A. J. Liu, and S. R. Nagel,
Phys. Rev. Lett.
 {\bf 95},
 098301
 (2005).

\bibitem{bri_dau_bir_bou_10}
C. Brito,   O. Dauchot,   G, Biroli, and   J.-P. Bouchaud,
Soft Matter
 {\bf 6},
 3013
 (2010).

\bibitem{owe_dan_13}
E. T. Owens and K. E. Daniels,
Soft Matter
 {\bf 9},
 1214
 (2013).

\bibitem{del_alt_kra_sca_14}
A. De Luca, B. L. Altshuler, V. E. Kravtsov, and A. Scardicchio,
Phys. Rev. Lett.
 {\bf 113},
 046806
 (2014).

\bibitem{tik_mir_16}
K. S. Tikhonov and  A. D. Mirlin,
Phys. Rev. B
 {\bf 94},
 184203
 (2016).

\bibitem{alt_iof_kra_16}
B. L. Altshuler, L. B. Ioffe, and V. E. Kravtsov,
arXiv:1610.00758.

\bibitem{bro_flo_fre_mel_pan_won_81}
T. A. Brody, J. Flores, J. B. French, P. A. Mello, A. Pandey, and S. S. M. Wong,
Rev. Mod. Phys.
 {\bf 53},
 385
 (1981).

\bibitem{eva_eco_90}
S. N. Evangelou and E. N. Economou,
Phys. Lett. A
 {\bf 151},
 345
 (1990).

\bibitem{bas_ale_alt_06}
D. M. Basko, I. L. Aleiner, and B. L. Altshuler,
Ann. Phys.
 {\bf 321},
 1126
 (2006).

\bibitem{oga_hus_07}
V. Oganesyan and D. A. Huse,
Phys. Rev. B
 {\bf 75},
 155111
 (2007).

%
%
\end{thebibliography}
\end{document}